\documentclass[conference]{IEEEtran}

\usepackage{cite}
\usepackage{amsmath,amssymb,amsfonts}
\usepackage{graphicx}
\usepackage{textcomp}
\usepackage{xcolor}
\usepackage{tabularx}
\usepackage{booktabs}
\usepackage{algpseudocode}

\begin{document}

\title{Musical Timing in Studio Recordings}

\author{
\IEEEauthorblockN{
Konstantinos Tsioutas and George Xylomenos}
\\
\IEEEauthorblockA{
Mobile Multimedia Laboratory, Department of Informatics,\\
School of Information Sciences and Technology, Athens University of Economics and Business, Greece}
}

\maketitle

\begin{abstract}
Studio recording techniques vary considerably, ranging from live recordings in a shared acoustic space, to isolated booth recording and overdubbing, where musicians record their parts separately, while listening to previously recorded material. These approaches differ in terms of physical co-presence, visual contact, sound leakage, and acoustic isolation, factors that may influence musical coordination. This study investigates whether the recording style affects timing precision; specifically, whether musicians performing together in the same space achieve tighter temporal coordination. To address these questions, two datasets with a total of 391 multi-track recordings were analyzed to measure the timing relationships between musicians. In addition, an automated sound-leakage detection method was employed to infer the recording conditions of each song: pairwise Mel-spectrogram similarities between the tracks were computed, and summary statistics from the resulting similarity matrices were used to identify recordings with shared acoustic content. The results suggest that recordings showing evidence of shared room acoustics and sound leakage tend to exhibit lower timing variability than highly isolated recordings, whereas overdubbing is associated with increased timing variability.
\end{abstract}


\section{Introduction}
\label{sec:intro}
In musical performance, timing accuracy describes how precisely the rhythmic onsets produced by musicians performing together are aligned in time. Timing accuracy strongly depends on the conditions of the musical performance: it depends on whether it is a live concert, a studio recording, or a musical rehearsal~\cite{art-of-studio}. In this study, we focus on the case of studio recordings and how musicians interact depending on the recording strategy used, focusing specifically on whether they are recording together or separately.

Recording studios and recording techniques have evolved in various ways over time. In early commercial recordings, musicians were directly recorded to a mastering medium that could not be modified using a single acoustic capturing device,  so all recordings were live, with musicians performing together, placed around the recording device. With the emergence of electronic analog recording equipment, musicians could be recorded using separate microphones, with a mixing console used to equalize their contributions; however, recording was still live to the mastering medium. As sound isolation between different microphones was not perfect, sound leakage was present in the recordings of each instrument. 

After the mid 1940s, the introduction of recording to magnetic tape eventually led to multitrack recording, where the microphone for each instrument was recorded to a separate magnetic track. This allowed audio tracks to be recorded at different points in time and overdubbed over the previously recorded tracks. The ability of tape to be erased and re-recorded individually for each track, allows combining basic tracks recorded with all musicians in the room with additional musical parts recorded later on by individual musicians. This removed the need of physical co-presence between musicians, allowing engineers to record one instrument at a time, with multiple takes per instrument, to allow the engineer to later choose the best take for the final mix. Sound leakage was thus only present in tracks recorded with multiple musicians in the same room.

In recent years, the move to digital recording rather than audio tapes and the emergence of \emph{Digital Audio Workstations}~(DAWs), has allowed recordings to use nearly infinite numbers of tracks, recorded either in unison or in isolation~\cite{pras}. Many recording studios offer multiple isolated recording booths, where acoustic audio sources can be recorded at the same time. In these booths, musicians can see each other via glass windows and listen to each other via closed type headphones. This allows a high degree of interaction, without sound leakage between tracks.

The research question that this study tries to explore is whether does the physical co-presence of musicians during recording affects timing. To examine this issue, we downloaded two datasets of studio multi-track recordings and analyzed their timing characteristics, as well as the similarities between the individual tracks of a recording using two different approaches, amplitude and spectral similarity. The results revealed that in recordings where we have evidence of physical co-presence, the variability of timing between musicians is significantly reduced, whereas in cases of possible overdubbing, timing variability increased.

The remainder of this paper is structured as follows. In Section~\ref{sec:evolution} we shortly present the evolution of studio recording techniques, while in Section~\ref{sec:timing_related}
we review musical timing related work. In Section~\ref{sec:timing_analysis} we present our dataset and its timing analysis, while in Section~\ref{sec:sound_leakage} we present our methodology for sound leakage detection. Section~\ref{sec:combination} we combine the results from the two previous sections, while in~\ref{sec:conclusions} we offer our concluding remaks and discuss future work.

\section{Evolution of  Studio Recording Techniques}
\label{sec:evolution}
To analyze timing in musical performance, it is critical to understand how a music song can be recorded and mixed. Understanding how the recording was made can help us understand the possible musical interactions between performers during the recording stage. These interactions can vary significantly depending on whether musicians played together at the same time, or played asynchronously, using overdubs. 

In the early recording era (late 1800s-1930s), performers played live to a single recording medium (initially a wax-coated cylinder and later a wax-coated disc), using a single acoustic capture device, called the acoustic horn. Balancing was achieved by the appropriate physical placement around the acoustic horn, without any possibility of later mixing or revision of the recordings. When electrical microphones came in to the studios and engineers could capture sound in more detail, using different microphones per instrument and consoles to adjust the microphone levels, recording remained for quite some time a one-shot process, as musicians were captured directly on a  write-once medium.

After the mid 1940s, magnetic tape revolutionized recording, allowing engineers to record multiple takes, edit, and assemble performances~\cite{zagorski}. Initially, corrections to a recording were made live (record something, play it back, record more on top), until multitrack tape emerged in the mid 1950s. Instead of capturing everything to one track, now each instrument or a few instruments could be captured to a separate track, allowing partial corrections to each track and final mixing at the end of the recording process. Studios started with 4-track, moving to 8, 16 and 24-track tape, offering many recording options. Of course, as microphones could not be perfectly isolated from other instruments, sound leakage between the tracks was common, despite the efforts of studio engineers to isolate the instruments with custom booths or sound panels.

The overdubbing technique, where a new recording is added to an existing recording, appeared quite early. Sidney Bechet, a jazz musician, Pierre Schaeffer, an electronic music composer, and Les Paul, an American jazz guitarist, used overdubbing to record  audio in the early 1930s. Les Paul however revolutionized multitrack recording after magnetic tape became commercially available, leading studios to embrace it. Overdubbing could be used to record multiple parts by the same musician, fix errors, or just allow tracks to be recorded by individual musicians in their own pace.

The emergence of digital audio recording, initially to digital audio tape and then to magnetic and solid state drives, was followed by the adoption of digital editing tools. Digital audio editing enabled recording and mixing very large numbers of tracks, as well as digital editing of tracks. Engineers could now record multiple takes of an instrument in parallel separate audio tracks; mistakes were acceptable, because they could be corrected in the editing stage, and recordings could be done asynchronously - for example, the drums of a song could be recorded in one day, and the bass in another day, with different takes spliced together or even adjusted to perfect timing.

Modern studios are equipped with advanced machinery that provides engineers with the flexibility to adopt hybrid techniques for recording. Live sessions can be combined with overdubs as needed. Studios are constructed with multiple recording booths and visual communication is feasible via large TV monitors and cameras, allowing perfect isolation between the instruments.

As the goal of this study is to compare the effects of different recording setups to timing, it is important to classify a production regarding the recording technique used. Unfortunately, the multitrack recordings that we have available are not fully documented in terms of their recording conditions. We can therefore only make educated guesses by analyzing the tracks themselves. Spefically, we are looking for audio leakage between tracks; this is an indication that the tracks were recorded together in the same space.

We can distinguish the following general recording categories: (a) live recording, where all musicians play together acoustic instruments, sharing the same acoustical room (with sound leakage), (b) live recording with electric and electronic instruments in the same room but with isolated audio recordings due to close-mics and/or direct to console recording using DI boxes (no sound leakage), (c) live recording, where all musicians play together but they are acoustically isolated in different booths (no sound leakage), (d) individual track recording (no sound leakage). For categories (a), (b) and (c) musicians have visual contact and especially for (a) and (b) they are placed in the same physical room. In category (d), musicians record their audio layer individually, listening to the prerecorded layer of another musician without any visual contact between them. Ideally, we would like to distinguish (d) from (a), (b) and (c), but lacking detailed recording information, we can use sound leakage to distinguish (a) from (b), (c) and (d).

\section{Related Work}\label{sec:timing_related}
Musical timing has been explored in many studies. 
Rasch \cite{10.1093/acprof:oso/9780198508465.003.0004} has conducted a pioneering effort in the field, introducing terms and concepts regarding synchronicity and coordination.
In the context of timing and synchronization, the author investigates \textit{asynchronization} $\Delta t$ of a pair of voices  which is defined as the time difference between two onsets of two different
musicians that should be played at the same time. It
can have a positive or  negative value, denoting the lead
or lag relationship between the two musicians. The standard deviation of $\Delta t$ denotes the variation in timing between musicians.
To evaluate $\Delta t$, Rasch asked two trios of musicians, one trio with oboe, clarinet and bassoon, and one trio with violin, viola and cello, to perform classical pieces. Performances were recorded with separated audio signals per instrument.  The mean value of $\Delta t$ is reported to be 36 ms.

Based on Rasch,  Clayton et. al. \cite{10.1525/mp.2020.38.2.136}
propose the \textit{Inter--Personal Musical Entrainment} (IME) which is formed by \textit{synchronization} and \textit{coordination}. The authors analyze recordings from six different music genres and compute timing metrics based on Rasch. The authors report that $\Delta t$ ranges between 15 and 35 ms but it varies depending on instrument type, event density, musical role, and metrical position. 

D' Amario et. al~\cite{10.3389/fpsyg.2018.01208} asked 22 singers to perform in duos in two conditions, either facing each other or facing in opposite directions. The authors report that the synchronization was much more precise and consistent when the singers were facing each other. They also compute the mean $\Delta t$ to be 59 ms and its standard deviation to be 60 ms.

Bishop~\cite{10.1177/1029864915570355}
investigates whether pianists rely on auditory cues or on visual cues during duet performances. Unlike D’Amario et al. (2018), this was not a live  experiment with the duo in the same room. Pianists played the secondo/accompaniment part while synchronizing with pre-recorded videos/audio of a primo performer, either a pianist or a violinist. The mean absolute of $\Delta t$ was measured in the range of 60--120 ms. The authors conclude that audio is the primary cue for duet synchronization, but visual cues become important at structurally difficult moments such as entrances, restarts, and long pauses.

Kawase et. al~\cite{10.3758/s13414-013-0568-0} investigate the role of eye gaze contact in piano duo performances. The authors report that piano duo performers look at each other mainly before difficult timing moments, and this mutual gaze -- especially together with visible movement -- helps them synchronize more accurately. They report that $\Delta t$ is large when performers cannot see each other, around 170--185 ms, and much smaller when visual information is available, around 47--62 ms. The best synchronization occurs when the performers can see each other’s movement, not just static gaze.

Finally, some studies have explored the role of visual contact in Networked Music Performance. In our previous work~\cite{tsioutas2026visual}, we experimented with 11 duos of musicians who performed on a local network under two basic conditions, with and without visual contact. The experiment was more complex, as the visual channel had a fixed delay of 32 ms while the delay in the audio channel  varied in the range from 12 to 72 ms one way. The basic finding of this study was that visual contact without synchronized video and audio strongly disrupted synchronization of musicians. Although the nature of a Networked Music Performance framework is much more complex than the conditions of a recording studio, the role of visual contact may expose similar effects in both cases.

\section{Multi-track Timing Analysis}
\label{sec:timing_analysis}

To investigate the $\Delta t$ between musicians we first need a dataset of multi-track recordings. One well-known multi-track dataset is MUSDB18~\cite{MUSDB18}, which is oriented to audio source separation. Unfortunately, MUSDB18  was not suitable for our analysis, because each stem is a product of audio separation of the master track. Instead, we chose two other datasets for timing analysis. The first dataset is the Cambridge dataset~\cite{380dataset} which contains 369 songs in multi-track format. The dataset covers a wide range of music genres, recorded in studios with unknown recording conditions. The second dataset is the Telefunken dataset~\cite{telefunken} which contains 22 songs recorded in the Telefunken studios in multi-track format. Unlike the Cambridge dataset, Telefunken provides some information through their website about each recording session and the techniques used.  

\begin{figure}[ht]
\centering
\includegraphics[width=0.9\columnwidth]{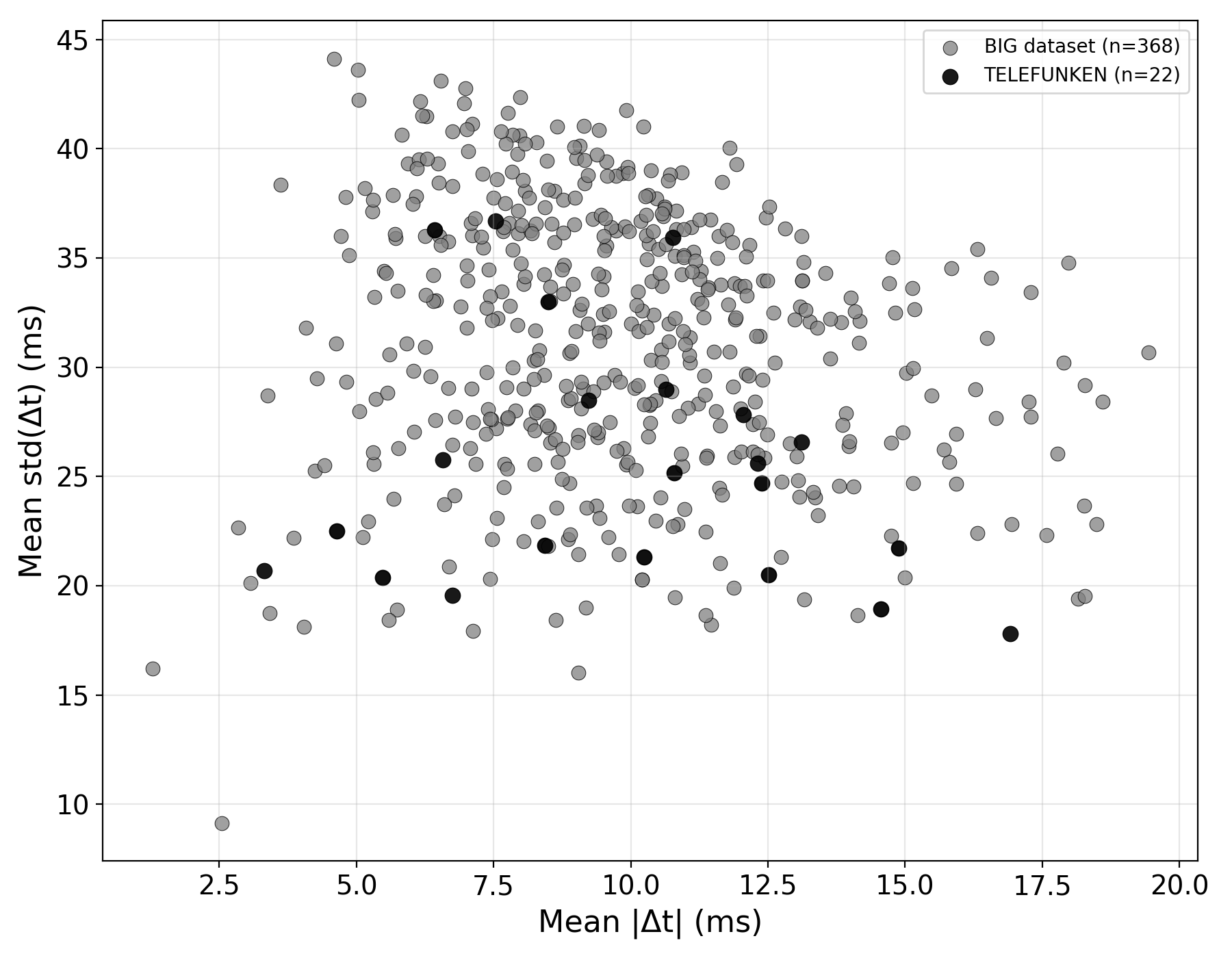}
\caption{Mean std($\Delta t$) against mean absolute $\Delta t$ distribution among both datasets}
\label{fig:timing_variablity}
\end{figure}

The core contribution of this study is the evaluation of the musical timing between musicians in studio recordings and its relationship with the recording setup. The core metric of our analysis is $\Delta t$~\cite{10.1093/acprof:oso/9780198508465.003.0004}.    
We performed a timing analysis to compute $\Delta t$ statistics (mean, median and standard deviation) for all stem--pairs per song, and per dataset. As shown in Figure~\ref{fig:timing_variablity} the mean absolute value of $\Delta t$ ranged from 2.5 to 18 ms while the standard deviation of $\Delta t$ ranged from 12 to 45 ms across all songs in both datasets. The graph reveals that there are songs with relatively low average $\Delta t$ but high std($\Delta t$) and vise versa. This means that despite leader--follower effects being low, timing instability is high and vise versa.  


A possible existence of separate clusters in this scatter plot, might reveal that the recording conditions can influence timing variability. 
In the next section, we will attempt to classify songs according to their recording conditions, to late correlate the recording conditions with the timing statistics.


\section{Sound Leakage Detection}
\label{sec:sound_leakage}

As mentioned above, our first (and largest) dataset does not offer any information about the recording setup, while the second (and smaller) dataset provides some. To search for evidence that musicians were recorded live, sharing the same acoustic space, a direction was to search for sound leaks between the audio stems of the recorded song. Sound leakage is the phenomenon where a microphone intended to capture an instrument also captures sounds from other instruments. Sound leakage can be easily detected by listening to an audio recording from a studio session, where instruments such as electric guitars and electric basses were performed in the same acoustic space with the drums. As electric guitars and basses are captured by microphones placed close to their amplifier cabins, sound leakage can occur very easily, as it is very hard to isolate the amplifiers without affecting the recorded audio. It is even harder to fully isolate drums, which are by their nature very loud, they take up considerable space and their sound heavily depends on room acoustics. This style of recording was very often adopted in cases of pop and rock bands, where the set included electric guitars, bass, pianos or organs with built-in amplifiers, drums and vocals. 

Sound isolation is the opposite of sound leakage, where audio files are totally clean from outside sounds and noises.  Sound isolation is the result of modern recording techniques in which overdubbing is used, with each track recorded separately and the musician performing against a recorded track.  
Since the sound leakage phenomenon is evidence of the possible co-presence of musicians in the same acoustic space, we focused our research in searching for methods for automatic sound leakage detection.  

One way for leakage detection is the comparison of two different stems of a multi-track session. The two stems can be compared regarding the similarity of their amplitudes or the similarity of their frequency spectra for the total duration of the song. If, for example, we consider two audio stems of a multi-track session, the comparison of their amplitude envelopes can result to a mean score close to 0, which means that the two stems are completely different in terms of amplitude over time, or to a score close to 1, which denotes that the two stems are almost indistinguishable.
If the stems are isolated clean recordings, the amplitude comparison is more likely to be close to 0, while if both stems are recordings of two different instruments, which were recorded in the same acoustic space with sound leakage, then the comparison of their amplitude envelopes can result in a relatively high score. These stems are possible candidates for sound leakage.

Although this simple comparison seems to solve the problem, it has limitations. If two isolated recordings of musicians playing the same melody are correlated regarding their amplitude similarity, then the result will also be close to 1, without any leakage; this is common with double tracking, dual guitarists playing the same part, and multi-tracked or backing vocals. Thus, high amplitude similarity does not necessarily mean that the cause is sound leakage.

Another option that can be used for stem correlation, is the Mel-Spectrogram. The Mel-Spectrogram is the spectral content of an audio file over time. Two different stems can be correlated regarding their spectral similarity and the score can be again in the range between 0 and 1.   
Mel-Spectrograms can also be used to detect sound leakage. A high spectral similarity score can denote the possible existence of sound leakage, although the same limitation exists with the amplitude correlation, albeit to a smaller extent. 

A second contribution of this study is the comparison of the two approaches, amplitude and spectral similarities, with respect to their usefulness in detecting sound leakage. In this direction, we computed metrics of both types for the entire dataset. First, we computed the \emph{Root Mean Square} (RMS) Similarity Matrix. The RMS is a measure of the energy of a signal over time. We first computed the RMS series of all stems of a song and then used the Pearson coefficient to compute the similarity between the RMS scores of each pair of stems. The RMS Similarity Score ranged from -1 to 1. The total matrix of RMS Similarities can be visualized using a heatmap graph.   
We applied these computations to two randomly chosen songs, one with no audible leakage 
and another with confirmed sound leakage, based on our own listening tests.

\begin{figure}[ht]
\centering
\includegraphics[width=0.95\columnwidth]{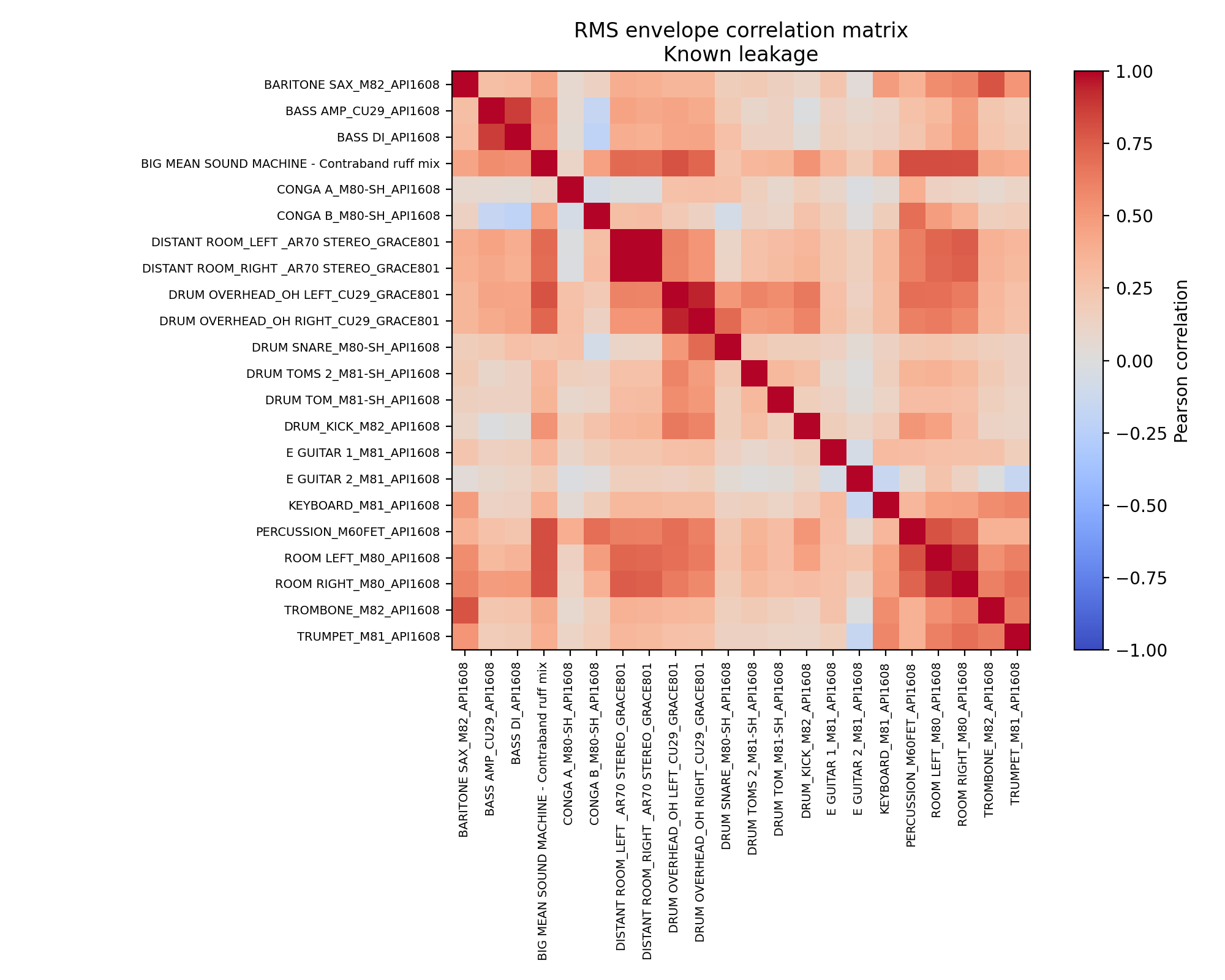}\caption{RMS envelope correlation matrix. Song with leakage}
\label{fig:rms_leakage}
\end{figure}
\begin{figure}[ht]
\centering
\includegraphics[width=0.9\columnwidth]{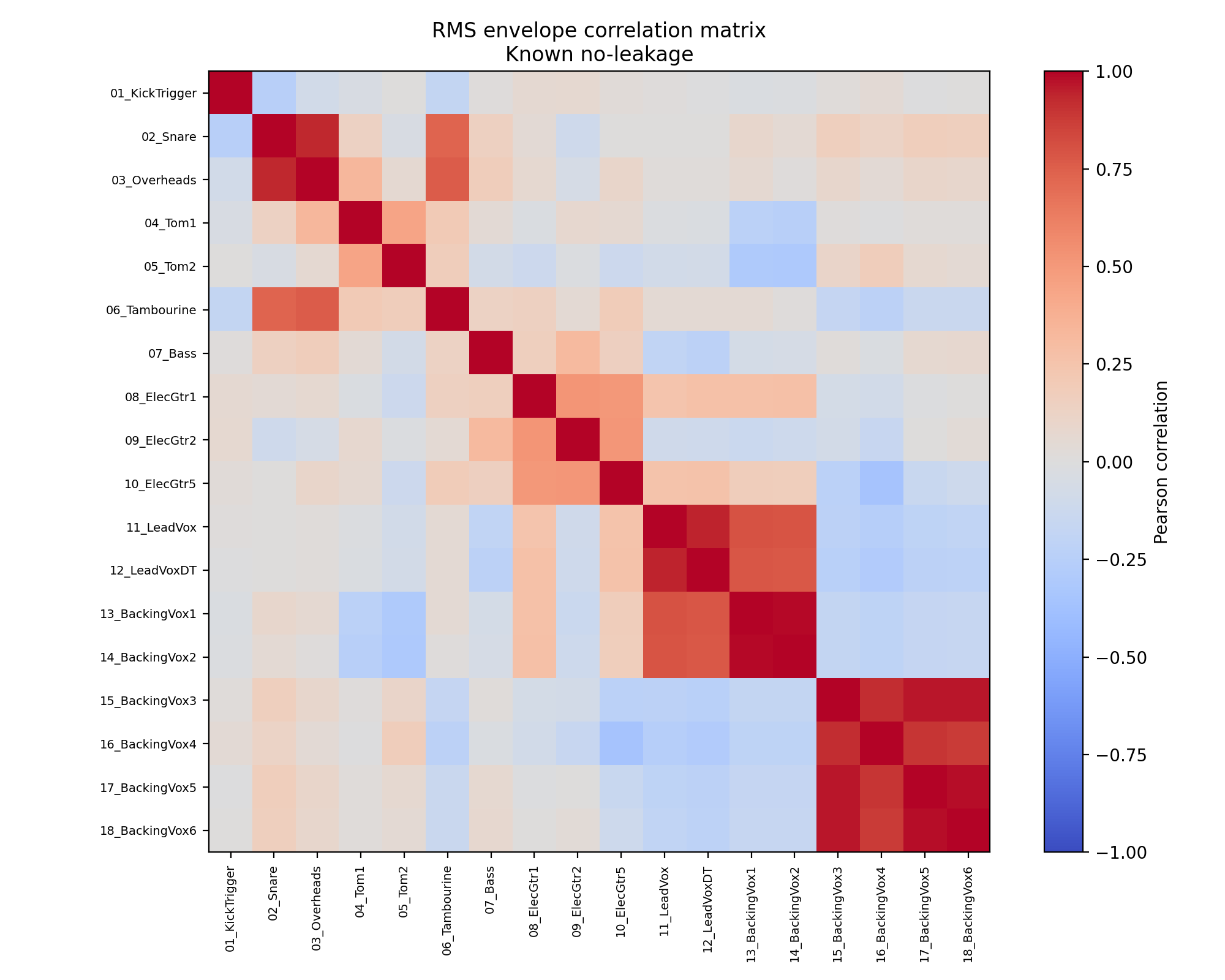}\caption{RMS envelope correlation matrix. Song without leakage}
\label{fig:rms_no_leakage}
\end{figure}
\begin{figure}[ht]
\centering
\includegraphics[width=0.95\columnwidth]{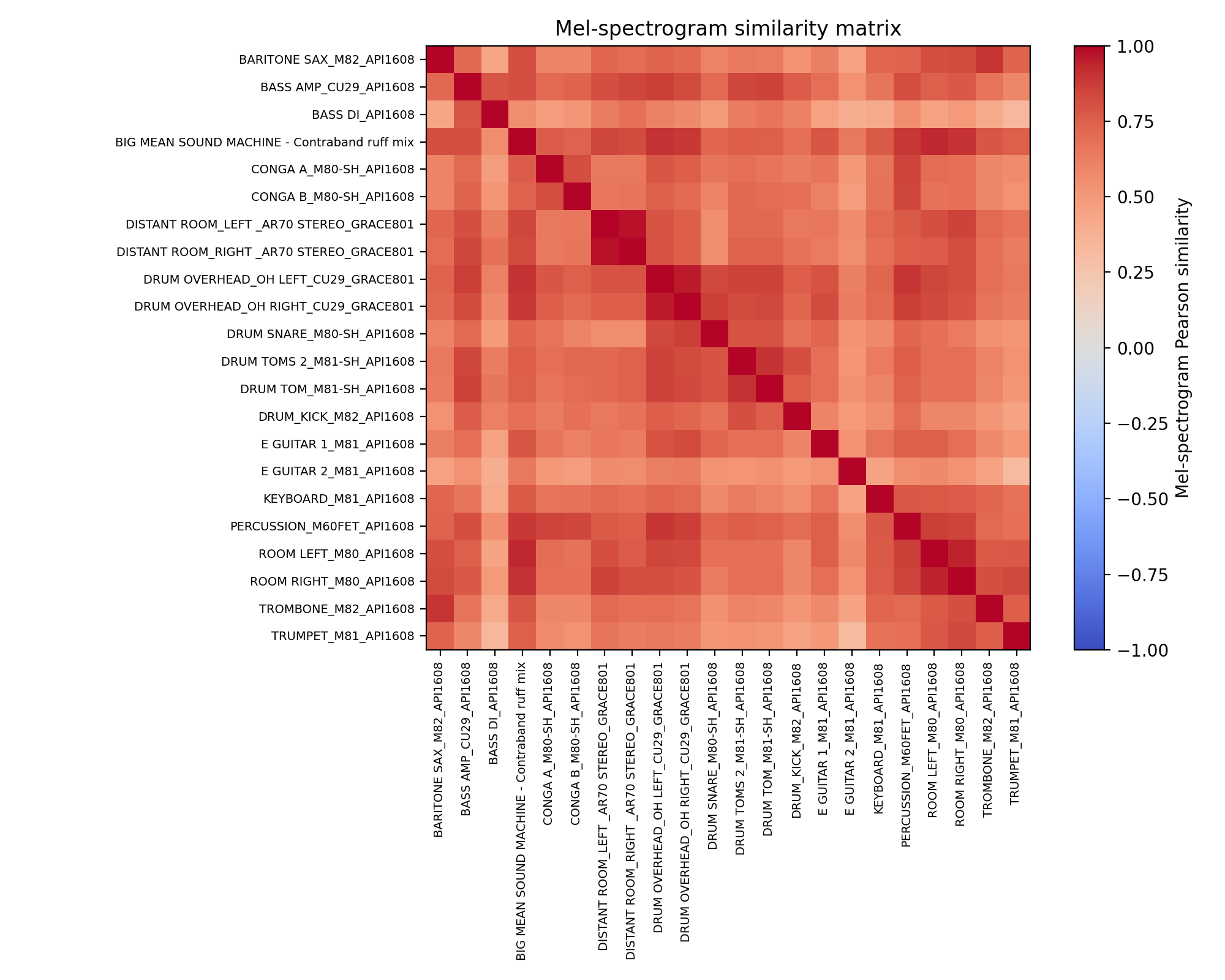}
\caption{Mel-Spectrogram correlation matrix. Song with leakage}
\label{fig:mel_leakage}
\end{figure}
\begin{figure}[ht]
\centering
\includegraphics[width=0.9\columnwidth]{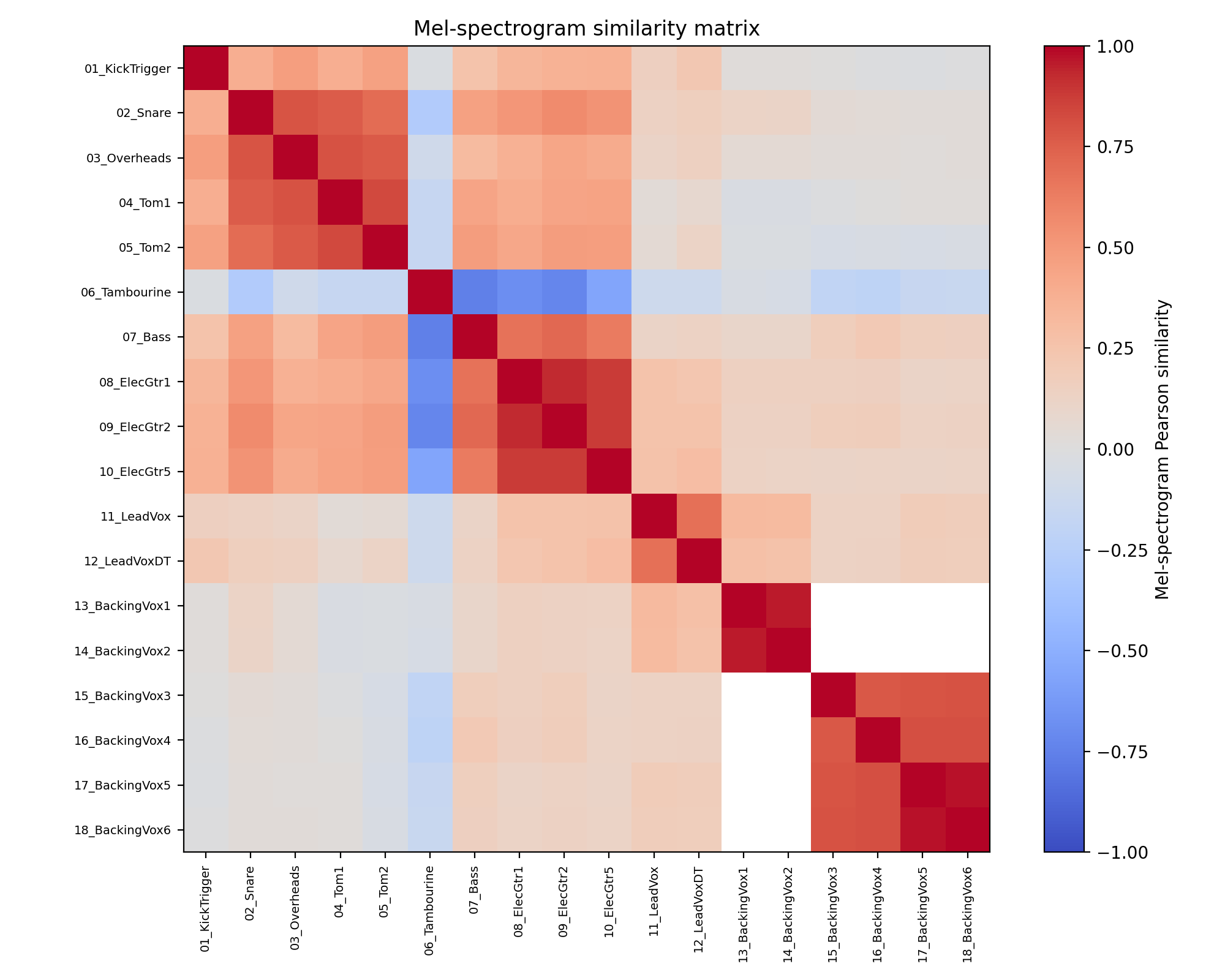}
\caption{Mel-Spectrogram correlation matrix. Song without leakage}
\label{fig:mel_no_leakage}
\end{figure}

We then computed the Mel-Spectrogram Similarity Matrix. The Mel-Spectrogram is a footprint of the spectrum of a stem over time and can also be computed with the librosa library~\cite{mcfee2015librosa}. For each song and each pair of stems, we computed the similarity of their Mel-Spectrograms. We applied these computations to the songs without leakage and the songs with leakage. The corresponding heatmaps are shown in Figures \ref{fig:rms_leakage} and \ref{fig:rms_no_leakage} (RMS envelope heatmap with/without leakage) and Figures \ref{fig:mel_leakage} and \ref{fig:mel_no_leakage} (Mel-Spectrogram heatmap with/without leakage).

Comparing Figures~\ref{fig:rms_leakage} and \ref{fig:rms_no_leakage} we can see that in a song with audible leakage, the RMS similarities between its stems are generally high (orange to red squares of the heatmap with score greater than 0.5) across all tracks, compared to a song without leakage where most of the squares are colored close to pink, white and sometimes blue. The diagonal represents each stem compared with itself, producing the highest score (perfect match). Note also the high similarities in the bottom red corner of the no leakage song, where the stems with vocals and backing vocals are shown; these are similar even though no leakage exists.

By comparing Figures~\ref{fig:mel_leakage} and \ref{fig:mel_no_leakage} where the Mel-Spectrogram similarities are shown, the same phenomenon is observed. In addition to the similarities between the vocal tracks (lower right), in Figure~\ref{fig:mel_no_leakage} we can also see similarities between the drum tracks (upper left) and the bass and electric guitars (middle of the diagram), as these instruments played the same chords at different octaves. In Figure~\ref{fig:mel_leakage} we get orange to red squares everywhere, indicating leakage.

Comparing the RMS and Mel-Spectrogram methods, it seems that the latter is more sensitive to leakage: while in the RMS method the song with leakage has more evidence of high correlation, with the Mel-Spectrogram the song with leakage is practically showing correlation across the entire set of tracks. However, the difference between songs with and without leakage is clear with both techniques.

\section{Influence of recording environment to timing} 
\label{sec:combination}

\begin{figure}[ht]
\centering
\includegraphics[width=0.9\columnwidth]{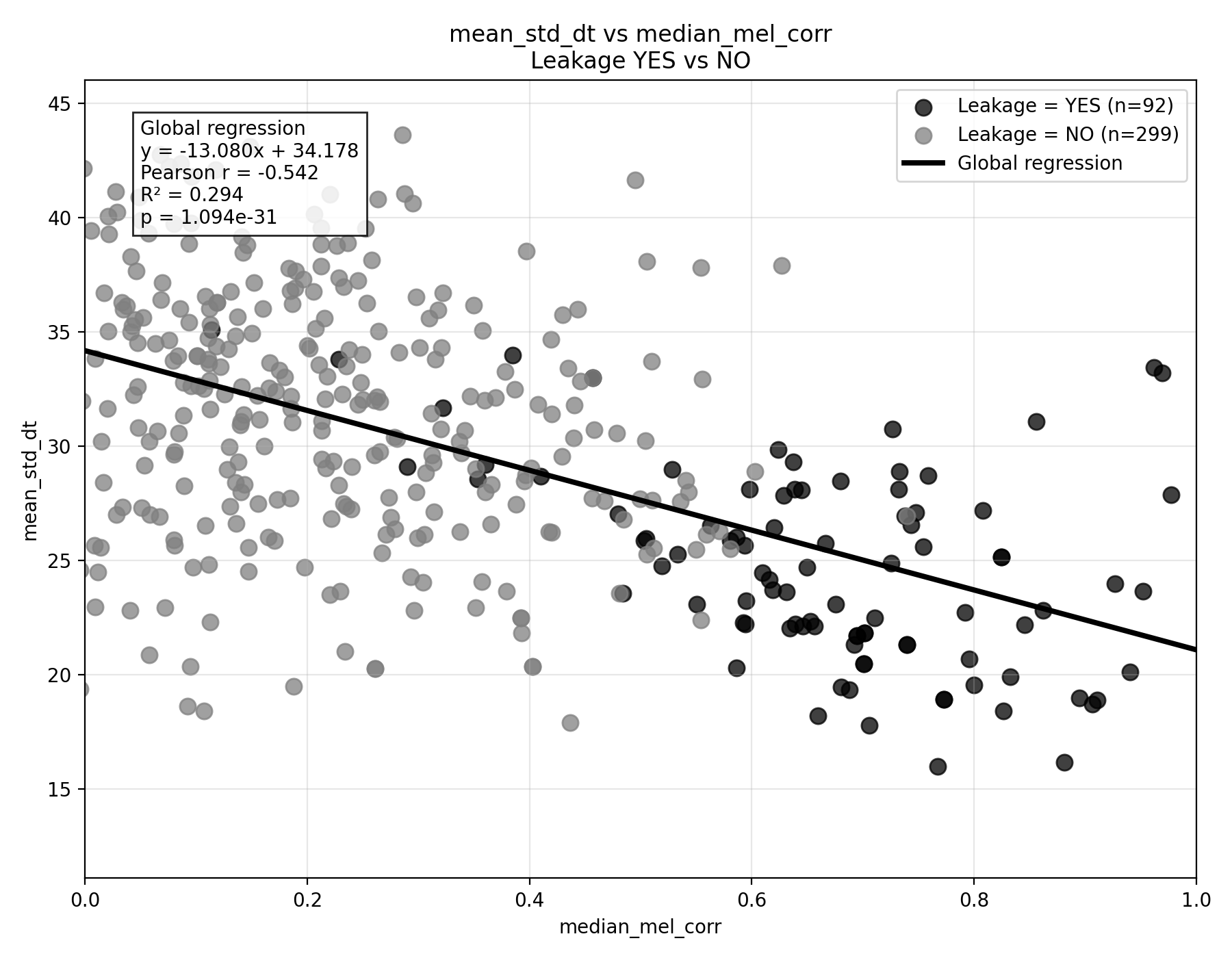}
\caption{Timing stability against median Mel-Spectrogram correlation score for both datasets, indicating for each data point whether leakage existed.}
\label{fig:timing_mel}
\end{figure}

To assess whether the recording environment has an effect on the timing statistics of the musicians, we combined the Mel-spectrogram similarity score computed in the previous section with the median value of the standard deviation computed in Section~\ref{sec:timing_analysis}; the results are shown is  Figure~\ref{fig:timing_mel}. The graph shows that the timing deviation is reduced as the Mel-Spectrogram similarity is increased, indicating that songs with higher spectral similarity between their onsets tend to show higher timing stability (lower median standard deviation of the $\Delta t$ metric). 

As there is no automated way to confirm that high spectral similarity was a result of sound leakage, we listened to a few random songs with high spectral similarity to verify whether it indicated leakage or not. By carefully listening to each stem of each of these songs, we gradually confirmed that high spectral similarity was indeed the result of sound leakage and not any other reason. 

As these few songs with high spectral similarity were confirmed to have sound leakage, we continue with the rest of the songs from the two datasets. By starting from songs with spectral similarity close to 1 and moving towards songs with lower similarity, a threshold of 0.45 was spotted under which almost all of the songs were found to be without leakage. The results of this process are shown in the two discrete clusters of Figure~\ref{fig:timing_mel} where the black dots are songs with confirmed sound leakage, while the light gray dots are songs with completely isolated stems.  

From the same figure it can be observed that the leakage = YES cluster spans into the leakage = NO cluster and vise versa. This phenomenon exists because many songs with high spectral similarity were found to be rhythmically tight, even though no sound leakage was detected in their stems. In contrast, songs with low spectral similarity  and more rhythmically loose, were found to include leakage. 

\begin{figure}[ht]
\centering
\includegraphics[width=0.95\columnwidth]{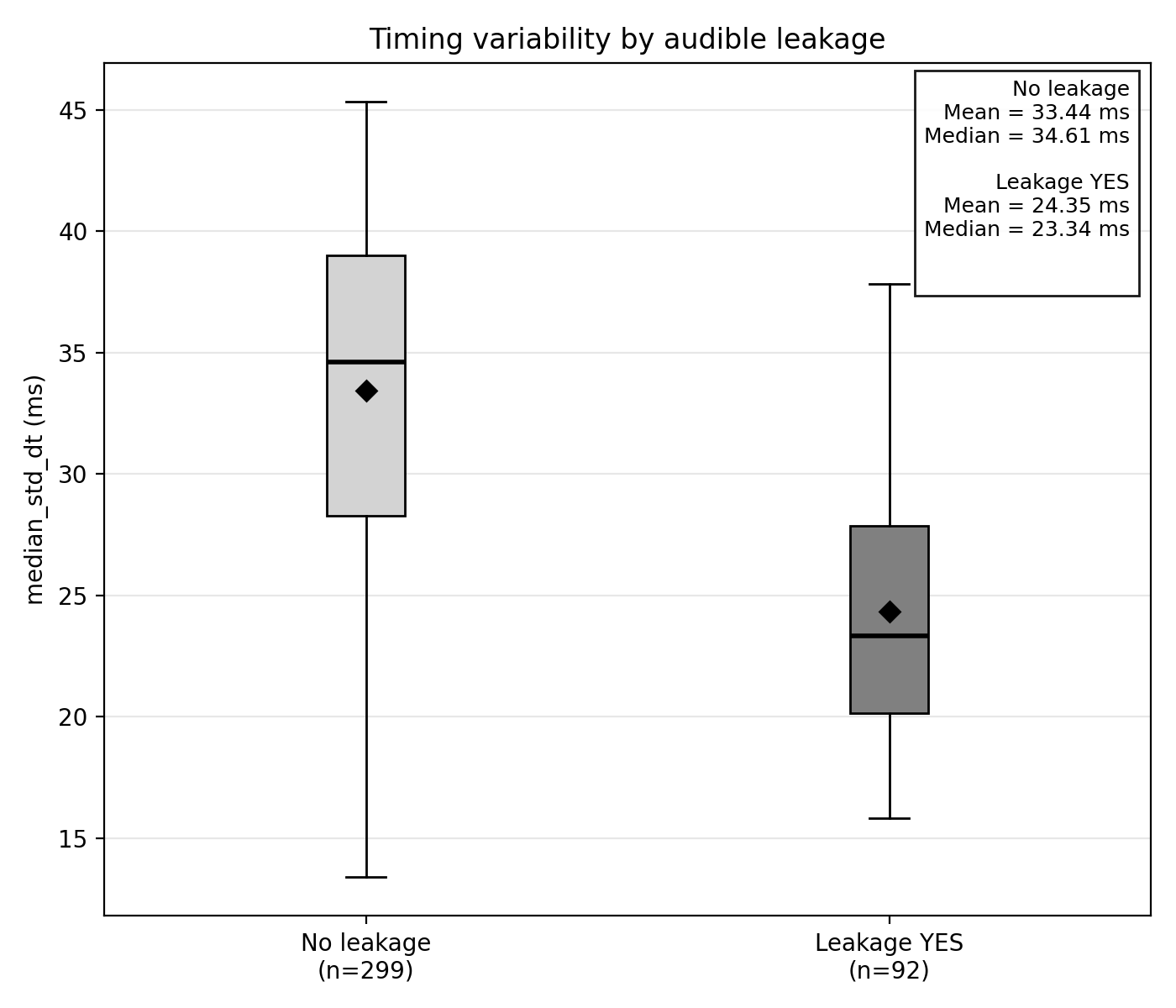}
\caption{Timing stability boxplot for each leakage condition.}
\label{fig:timing_leakage}
\end{figure}

In Figure~\ref{fig:timing_leakage} the variation of $\Delta t$ against the leakage condition is shown for 392 songs using a separate boxplot for each leakage condition (leakage or no leakage). We can observe that in songs with leakage the median value of rhythmic variation is 24.35 ms, while in the boxplot without leakage this value is 34.61, almost 10 ms higher, confirming the significant effect of sound leakage ($p < 0.05$).


To evaluate whether Mel-spectrogram correlation provided a more effective indicator of audible  leakage than RMS correlation, both measures were compared against the leakage condition. The median Mel correlation exhibited substantially larger effect sizes (Cohen's d = 2.78, Cliff's $\delta$ = 0.92) than the median RMS correlation (Cohen's d = 1.61, Cliff's $\delta$ = 0.74), indicating better discrimination between leakage-positive and leakage-negative recordings. Cohen's d was computed as the difference between the means of the leakage-positive and leakage-negative groups divided by their pooled standard deviation, providing a standardized measure of effect size independent of the scale of the correlation metric. We therefore propose using the Mel-spectrogram method for future studies.

While these results support the hypothesis that when recording together musicians exhibit better timing synchronization, we must note again that sound leakage is an imperfect indicator of how a song was recorded. Specifically, a song where sound leakage exists, is most likely to have been recorded with musicians together in a room. However, the absence of sound leakage does not necessarily indicate that the musicians performed separately, as live performances with good audio isolation or with instruments that do not need direct microphones will not lead to sound leakage. In addition, the detection of sound leakage with automated means, whether with Mel-spectrograms or with RMS correlation, is also imperfect, as noted above.

\section{Conclusions}
\label{sec:conclusions}
In this study, we examine the effects of sound leakage on timing stability between musicians in studio recordings. Two datasets of 391 songs in total were examined with respect to timing precision and the existence of sound leakage. The latter was examined and confirmed by listening to all songs manually. We perform spectral similarity computations as a tool for sound leak detection, finding that high spectral similarity between song stems is strongly correlated with sound leakage and low timing variation. This is strong evidence that co-presence between musicians leads to lower timing variations in their ensemble performances.

Due to the imperfect correspondence between sound leakage and recording conditions, future work will focus on locating and analyzing multi-track recordings with detailed provenance. Another, more costly, option is to perform experiments with the specific purpose of comparing multi-track recordings with the same musicians recording either together or in isolation, to allow attributing any differences between the performances to the actual recording setting rather than differences between musicians.

\bibliographystyle{IEEEtran}
\bibliography{bibtex}

@String{Academic = "Academic Press" }

@incollection{10.1093/acprof:oso/9780198508465.003.0004,
    author = {Rasch, Rudolf A.},
    isbn = {9780198508465},
    title = {Timing and synchronization in ensemble performance},
    booktitle = {Generative Processes in Music: The Psychology of Performance, Improvisation, and Composition},
    publisher = {Oxford University Press},
    year = {2001},
    month = {01},
    doi = {10.1093/acprof:oso/9780198508465.003.0004},
    url = {https://doi.org/10.1093/acprof:oso/9780198508465.003.0004},
    eprint = {https://academic.oup.com/book/0/chapter/154427776/chapter-ag-pdf/44959937/book_8555_section_154427776.ag.pdf},
}

@article{10.1525/mp.2020.38.2.136,
    author = {Clayton, M. and Jakubowski, K. and Eerola, T. and Keller, P. E. and Camurri, A. and Volpe, G. and Alborno, P.},
    title =  {Interpersonal entrainment in music performance: Theory, method and model.},
    journal = {Music Perception: An Interdisciplinary Journal, 38(2), 136-194.},
    pages = {136-194},
    volume = {38(2)},
    doi ={https://doi.org/10.1525/mp.2020.38.2.136},
    year = 2020
}

@misc{MUSDB18,
  author = {Rafii, Zafar and
                  Liutkus, Antoine and
                  Fabian-Robert St{\"o}ter and
                  Mimilakis, Stylianos Ioannis and
                  Bittner, Rachel},
  title = {The {MUSDB18} corpus for music separation},
  month = dec,
  year  = 2017,
  doi   = {10.5281/zenodo.1117372},
  url   = {https://doi.org/10.5281/zenodo.1117372}
}

@misc{art-of-studio,
    author = {Williams, Allan },
    title = {Divide and Conquer: Power, Role Formation, and Conflict in Recording Studio Architecture},
    url = {https://www.arpjournal.com/asarpwp/divide-and-conquer-power-role-formation-and-conflict-in-recording-studio-architecture/}
    
}

@article{pras,
author = {Pras, Amandine and Guastavino, Catherine and Lavoie, Maryse},
year = {2013},
month = {03},
pages = {612–626},
title = {The impact of technological advances on recording studio practices},
volume = {64},
journal = {Journal of the American Society for Information Science and Technology},
doi = {10.1002/asi.22840}
}

@book{zagorski,
    editor = {Zagorski-Thomas, S. and Isakoff, K. and Lacasse, S. and Stévance},
    title ={The Art of Record Production: An Introductory Reader for a New Academic Field (1st ed.)} ,
    publisher = {Routledge},
    year = {2012},
    url = {https://doi.org/10.4324/9781315612638}
}

@misc{380dataset,
    title = {Mixing Secrets For The Small Studio  Additional Resources},
    url ={https://www.cambridge-mt.com/ms3/mtk/} 
    
}

@misc{telefunken,
    title = {Telefunken Elektrokustik
Multi-track Sessions},
    url = {https://www.telefunken-elektroakustik.com/multitracks/} 
}

@ARTICLE{10.3389/fpsyg.2018.01208,
    
AUTHOR={D'Amario, Sara  and Daffern, Helena  and Bailes, Freya },
           
TITLE={Synchronization in Singing Duo Performances: The Roles of Visual Contact and Leadership Instruction},
          
JOURNAL={Frontiers in Psychology},
          
VOLUME={9},
  
YEAR={2018},
  
 
DOI={10.3389/fpsyg.2018.01208},
  
ISSN={1664-1078}
  
}

@article{10.1177/1029864915570355,
    author ={Bishop, Laura and  Goebl, Werner},
    title ={When they listen and when they watch: Pianists' use of nonverbal audio and visual cues during duet performance.}  ,
    journal = {The journal of the European Society for the Cognitive Sciences of Music},
    year = {2015},
    volume = {19}
}

@article{10.3758/s13414-013-0568-0,
author = {Kawase, Satoshi},
year = {2013},
month = {10},
pages = {},
title = {Gazing behavior and coordination during piano duo performance},
volume = {76},
journal = {Attention, perception \& psychophysics},
doi = {10.3758/s13414-013-0568-0}
}

@inproceedings{tsioutas2026visual,
  author    = {Tsioutas, Konstantinos and Xylomenos, George and Alexandraki, Chrisoula},
  title     = {Visual Contact in Networked Music Performance},
  booktitle = {Proceedings of the 32nd International Conference on Telecommunications (ICT)},
  year      = {2026}
}

@inproceedings{mcfee2015librosa,
  author    = {McFee, Brian and Raffel, Colin and Liang, Dawen and Ellis, Daniel P. W. and McVicar, Matt and Battenberg, Eric and Nieto, Oriol},
  title     = {librosa: Audio and Music Signal Analysis in Python},
  booktitle = {Proceedings of the 14th Python in Science Conference},
  pages     = {18--25},
  year      = {2015},
  doi       = {10.25080/Majora-7b98e3ed-003}
}

\end{document}